\documentclass[%
 aip,
 jmp,%
 amsmath,amssymb,
 reprint,%
 preprintnumbers,
]{revtex4-1}

\usepackage{graphicx}
\usepackage{dcolumn}
\usepackage{bm}

\begin{document}


\title{Technique for high axial shielding factor performance of
large-scale, thin, open-ended, cylindrical Metglas magnetic shields}

\author{S.\ Malkowski}
\author{R.\ Adhikari}
\author{B.\ Hona}
\author{C.\ Mattie}
\author{D.\ Woods}
\author{H.\ Yan}
\author{B.\ Plaster}
\affiliation{Department of Physics and Astronomy, University of Kentucky,
Lexington, Kentucky 40506, USA}

\date{\today}

\begin{abstract}
Metglas 2705M is a low-cost commercially-available, high-permeability
Cobalt-based magnetic alloy, provided as a 5.08-cm wide and
20.3-$\mu$m thick ribbon foil.  We present an optimized construction
technique for single-shell, large-scale (human-size), thin, open-ended
cylindrical Metglas magnetic shields.  The measured DC axial and
transverse magnetic shielding factors of our 0.61-m diameter and
1.83-m long shields in the Earth's magnetic field were 267 and 1500,
for material thicknesses of only 122 $\mu$m (i.e., 6 foil layers).
The axial shielding performance of our single-shell Metglas magnetic
shields, obtained without the use of magnetic shaking techniques, is
comparable to the performance of significantly thicker,
multiple-shell, open-ended Metglas magnetic shields in
comparable-magnitude, low-frequency applied external fields reported
previously in the literature.
\end{abstract}



\maketitle

\section{Introduction}
The suitability of Metglas 2705M for the construction of small- and
large-scale (human-size) magnetic shields has been discussed
extensively in the literature \cite{sasada88, sasada96, sasada96b,
sasada00, nagashima02}.  This commercially-available \cite{metglas}
amorphous, Cobalt-based magnetic alloy is provided as a 5.08-cm wide
and 20.3-$\mu$m thick ribbon foil, at a relatively low cost of
665 USD per kilogram (yielding 113 m of material).  In addition
to its relatively low cost (as compared, for example, to standard
$\mu$-metal magnetic shields), there are several advantages to the
construction of magnetic shields with Metglas, including: (a) the
material's high permeability (previous studies \cite{sasada88}
determined the permeability to be $\sim 5 \times 10^5$ under magnetic
shaking conditions); (b) the amorphous nature of the material,
permitting construction of magnetic shields of nearly any geometric
shape; and (c) the ability to re-use the foils for different magnetic
shield assemblies, thereby reducing costs.

In this article we present results from optimization studies of
construction techniques for large-scale (human-size) Metglas magnetic
shields, with diameters of 0.61 m and lengths of 1.83 m.  Our
measurements of the DC axial and transverse shielding factors were
carried out in the Earth's magnetic field, without the use of magnetic
shaking techniques, as employed in previous studies of Metglas
magnetic shields \cite{sasada96, sasada96b, sasada00, nagashima02}.
Our results for the axial shielding factors of our thin, single-shell
assembly in the Earth's DC magnetic field are comparable to the axial
shielding factors reported previously of a significantly thicker,
multiple-shell assembly that were obtained in low-frequency, applied
external fields of magnitude comparable to the Earth's field
\cite{sasada88, sasada96, sasada96b, sasada00, nagashima02}.

\section{Technique}
\begin{figure}[t]
\begin{center}
\includegraphics[angle=270,scale=0.36,clip=]{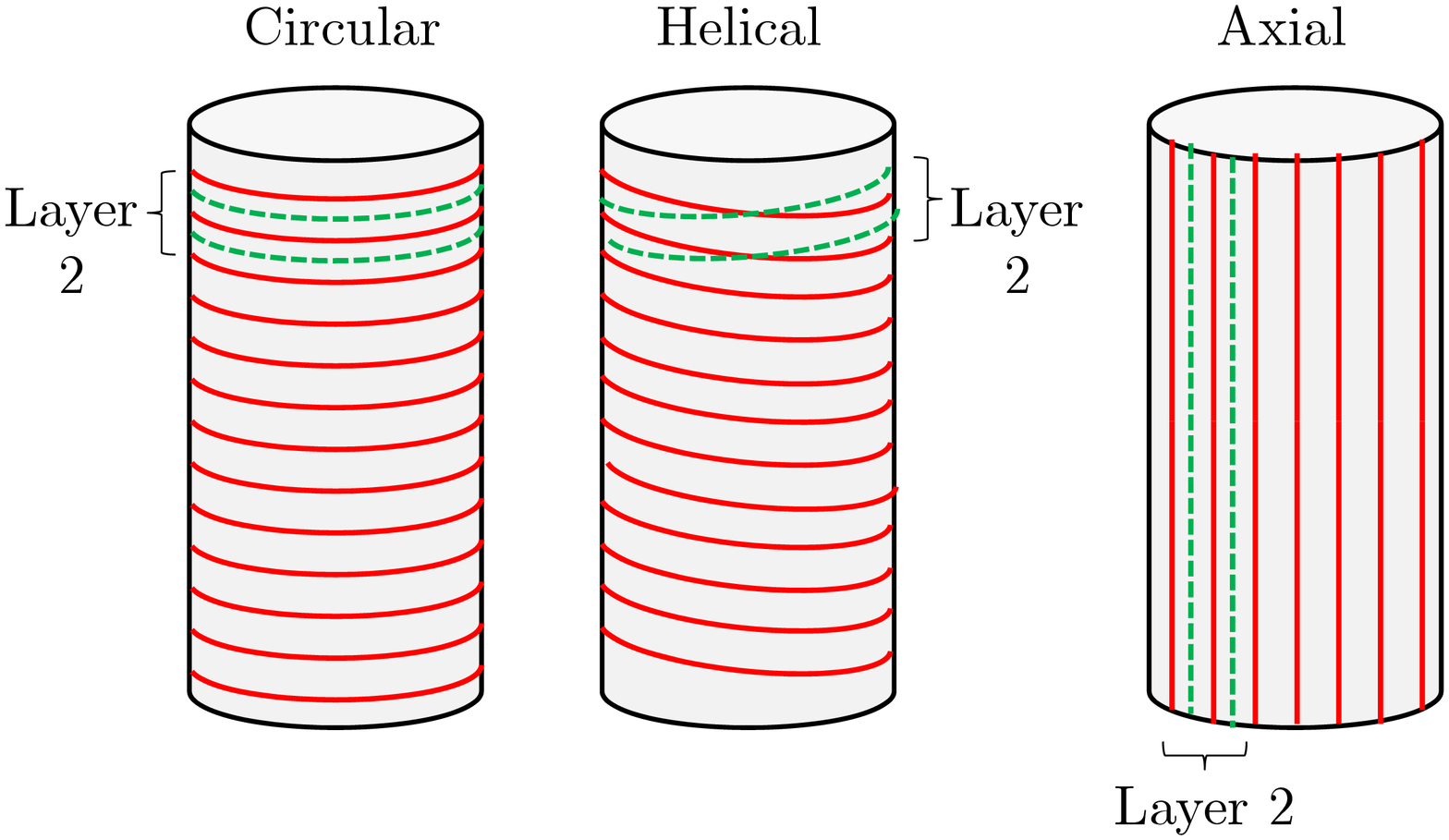}
\caption{(Color online) Illustration of the different Metglas winding
techniques: circular, helical, and axial.  The dashed lines labeled
``Layer 2'' indicate the placement and/or helicity of the second layer
relative to the underlying first layer.  See text for details.}
\label{fig:metglas_winding_techniques}
\end{center}
\end{figure}

We constructed our Metglas magnetic shields by ``winding'' the foil
onto the surface of cylindrical cardboard support forms.  We tested
several different winding techniques, which are illustrated
schematically in Fig.\ \ref{fig:metglas_winding_techniques}.

\begin{itemize}

\item
``Circular'' windings: Here, the foils were cut to a length equal to
(actually, slightly greater than) the circumference of the shield, and
then wound onto the surface of the form circumferentially.  Thin
(25.4-$\mu$m thick) Kapton \cite{kapton} tape was used to secure one
end of these windings onto the form surface, and to secure the other
(slightly overlapping) end to the short overlapped section of foil.
The first layer of the shield consisted of $N$ of these such circular
windings, with $N$ equal to the shield length divided by the foil
width of 5.08 cm.  Note that we were careful to minimize the gaps
between adjacent circular windings.

The second layer of the shield was then wound in exactly the same
manner, but an important point is that this layer consisted of $N-1$
circular windings, with the positions of the circular windings along
the shield axis offset from those in the first layer by one-half of
the foil width (i.e., by 2.54 cm) in order to ``cover'' the gaps
between the adjacent circular windings in the first layer (see Fig.\
\ref{fig:metglas_winding_techniques}).  The third/fifth/etc.\ and
fourth/sixth/etc.\ layers were then constructed identically to the
first and second layers, respectively.

\item
``Helical'' windings: Here, we used Kapton tape to adhere one end of
the foil onto the form surface, and then proceeded to wind the foil as
a single, continuous strip onto the surface of the shield from one end
to the other in a spiraling, helical manner.  Each successive turn
overlapped the previous by approximately one-half of the material
width, and an important point is that Kapton tape was used to secure
each successive turn to the previous turn.  Thus, a single-layer
helical winding yielded an effective material thickness of $\sim 1.5$
times that of a single-layer circular winding (which, correspondingly,
required $\sim 1.5$ times more material).  Successive layers were
wound with opposite helicities (e.g., first layer was wound as a
right-handed helix, second layer as a left-handed helix, etc.).

\item
``Axial'' windings: Here, the foils were cut to a length equal to the
length of the shield, and then aligned along the form surface in
the axial direction.  Again, Kapton tape was used to secure the two
ends of the foil to the form surface.  The first layer then consisted
of $M$ of these such axial windings, with $M$ equal to the shield
circumference divided by the foil width of 5.08 cm.  Again, we were
careful to minimize the gaps between adjacent axial windings.

The second layer of an axial winding was then constructed in exactly
the same manner as the first layer.  However, just as with the
circular windings, the positions of the foils along the circumference
were offset from those in the first layer by one-half of the foil
width in order to ``cover'' the gaps between the adjacent foils (see
Fig.\ \ref{fig:metglas_winding_techniques}).  Note that the second
layer also required $M$ foils.  The third/fifth/etc.\ and
fourth/sixth/etc.\ layers were then constructed identically to the
first and second layers, respectively.

\end{itemize}

We believe a final important feature of our construction technique is
that there were no (intentional) gaps between successive layers (e.g.,
between the first and second layers); the foils comprising any two
successive layers were in direct contact.  Also, if the shield
included one or more axial foils, the outermost layer was then covered
with a layer of plastic shrink wrap, which served to smooth the axial
foils onto the curved shape of the cylindrical form surface.  Finally,
degaussing coils were wound onto the form in a toroidal geometry for
circular and helical windings and in circular geometries (at multiple
positions along the shield axis) for axial windings.

\section{Results}

\begin{figure}[t]
\begin{center}
\includegraphics[scale=0.50]{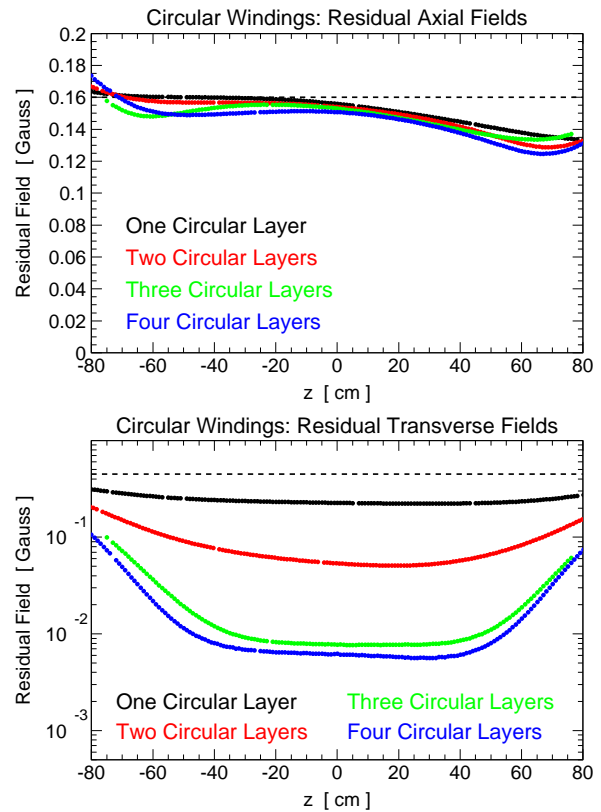}
\caption{(Color online) Measured residual axial (top panel) and
transverse (bottom panel) fields within a Metglas shield consisting
only of circular windings, for successive increased layering.  Note
that additional layering resulted in only marginal improvement to the
axial shielding.  The dashed lines indicate the external background
fields.}
\label{fig:metglas_windings_publication_1}
\end{center}
\end{figure}

\begin{figure}[t]
\begin{center}
\includegraphics[scale=0.50]{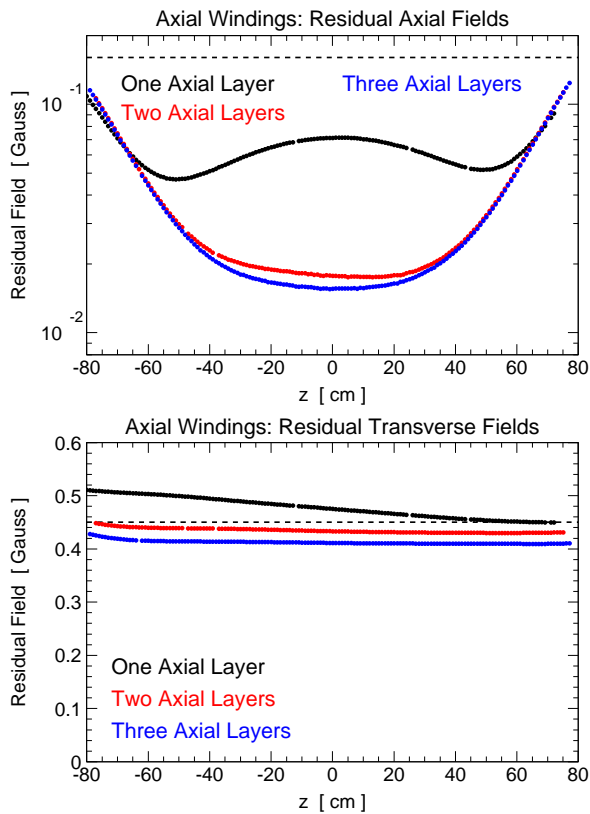}
\caption{(Color online) Measured residual axial (top panel) and
transverse (bottom panel) fields within a Metglas shield consisting
only of axial windings, for successive increased layering.  Note that
additional layering resulted in only marginal improvement to the
transverse shielding.  The dashed lines indicate the external
background fields.}
\label{fig:metglas_windings_publication_2}
\end{center}
\end{figure}

All of our measurements of the residual shielded fields within our
magnetic shields were performed with an automated magnetic mapping
system which consisted of a computer-controlled, three-axis stepper
motor assembly.  This system controlled the movement of a low-noise
triple-axis fluxgate magnetometer (with a resolution better than $\pm
10$ $\mu$Gauss), which was mounted on the end of a 2.1-m long
non-magnetic arm (made of G10).  Measurements of the residual shielded
fields in the Earth's magnetic field were conducted after a 60 Hz AC
degaussing cycle.  Our primary results are as follows.

First, as briefly noted in Ref.\ 1, for cylindrical Metglas magnetic
shields, axially-oriented (transversely-oriented) foils are not
effective at shielding transverse (axial) external fields.  However,
explicit data were not shown.  We present data demonstrating this
effect in Figs.\ \ref{fig:metglas_windings_publication_1} and
\ref{fig:metglas_windings_publication_2}, which show results from
measurements of the residual axial and transverse fields along the
axis of magnetic shields consisting only of circular or axial
windings.  (Results from shields consisting only of helical windings
are similar to those consisting only of circular windings.)  These
data show conclusively that circular windings are effective at
shielding transverse external fields, but provide very little
shielding against axial external fields, even with additional
layering.  Similarly, axial windings are effective at shielding
axial external fields, but provide essentially no shielding against
transverse external fields.

\begin{figure}[t]
\begin{center}
\includegraphics[scale=0.50]{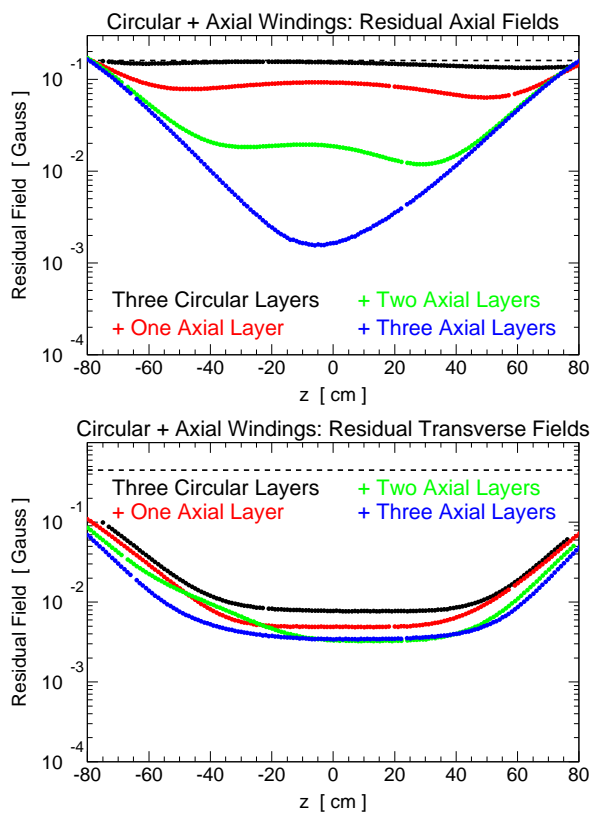}
\caption{(Color online) Measured residual axial (top panel) and
transverse (bottom panel) fields within a Metglas shield with a
circular plus axial winding combination, for successive increased
layering.  The dashed lines indicate the external background fields.}
\label{fig:metglas_windings_publication_4}
\end{center}
\end{figure}

\begin{figure}[t]
\begin{center}
\includegraphics[scale=0.50]{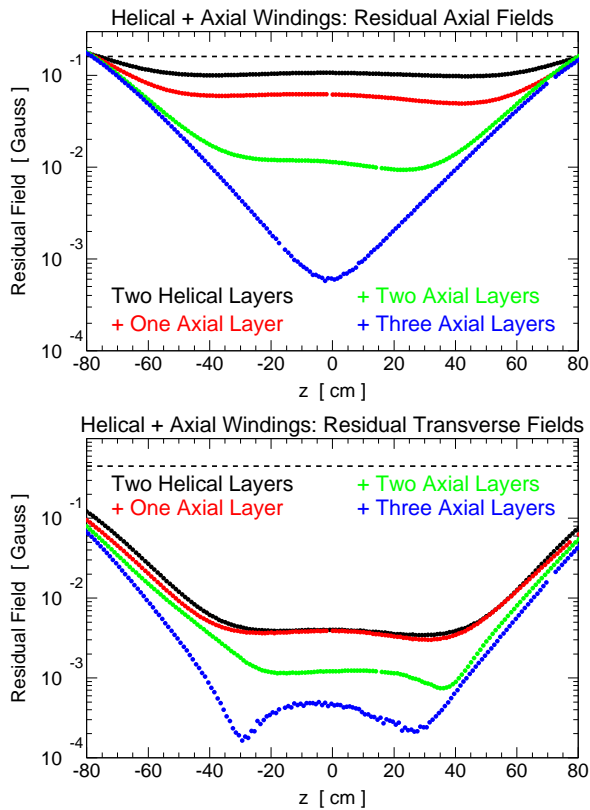}
\caption{(Color online) Measured residual axial (top panel) and
transverse (bottom panel) fields within a Metglas shield with a
helical plus axial winding combination, for successive increased
layering.  The dashed lines indicate the external background fields.}
\label{fig:metglas_windings_publication_5}
\end{center}
\end{figure}

Second, Figs.\ \ref{fig:metglas_windings_publication_4} and
\ref{fig:metglas_windings_publication_5} show results from
measurements of the residual axial and transverse fields within
shields which consisted of a circular plus axial winding combination,
and a helical plus axial winding combination.  We note again that
these (open-ended) single-shell shields were 0.61~m in diameter, and
1.83~m in length, and consisted of only 5--6 layers of Metglas (i.e.,
thicknesses of 102--122 $\mu$m).  The results shown there were
obtained with the axial windings wound directly onto (i.e., on top of,
and in direct contact with) the underlying circular or helical
windings.  [Note that similar results were obtained with alternating
layers of circular and axial windings.]  As can be seen there,
slightly better results were obtained for the helical plus axial
winding combination, as compared to the circular plus axial winding
combination (for similar material thicknesses).

For our helical plus axial winding combination, the residual axial and
transverse fields in the center of our shield were $\sim 600$
$\mu$Gauss and $\sim 200$--400 $\mu$Gauss, for external axial and
transverse background fields of 0.16 Gauss and 0.45 Gauss,
respectively.  Thus, our measured axial and transverse shielding
factors were $S_A = 267$ and $S_T = 1500$.  Using standard formulae
for the transverse \cite{sumner87} and axial \cite{edm_book} shielding
factors of a single-shell cylindrical shield with radius $R$ ($=
0.305$ m), length $L$ ($= 1.83$ m), thickness $t$ ($= 122$ $\mu$m),
and relative permeability $\mu$,
\begin{equation}
S_T = \frac{\mu t}{2R},~~~~~~~~~~
S_A \approx \frac{2\mu t R^{1/2}}{L^{3/2}},
\end{equation}
our results suggest a relative permeability of $\mu \sim 6 \times
10^6$ in these external background fields, with consistent results
obtained from the axial ($5.0 \times 10^6$) and transverse
($7.5 \times 10^6$) shielding factor measurements.  Note that the
values of these permeabilities are consistent, even though the above
formula for the axial shielding factor assumes a cylindrical shield
with end caps, whereas our shields were open-ended.

It is interesting to note that the axial shielding performance of our
thin, single-shell shield in the Earth's DC magnetic field is comparable
to that achieved in Ref.\ 4 with a four-shell nested shield assembly
in a low-frequency 0.1 G external field; this multiple-shell assembly
included a single Permalloy shield and a three-shell Metglas shield
assembly (composed of a total of 78 Metglas foil layers) with
dimensions slightly larger than ours: 0.7-m diameters and 2.7-m
lengths.  Axial and transverse shielding factors of 180 and 5000 were
achieved without magnetic shaking, with axial shielding factors of
$\sim 15$ reported for single-shell Metglas shields (consisting of up
to 30 Metglas foil layers).

Note that the measured residual fields near the ends of our cylinders
were, in some cases, larger than the external background fields (see,
e.g., Figs. \ref{fig:metglas_windings_publication_1} and
\ref{fig:metglas_windings_publication_2}).  We attribute this to the
magnetization of the shield in the external background field coupled
to a theoretical residual field profile, especially for the residual
axial (transverse) fields when the Metglas winding was circular
(axial) [i.e., for the winding scenario demonstrated to yield only
marginal shielding for a particlar orientation].  This effect was
observed in finite-element-analysis calculations.

\section{Summary}
In summary, we have demonstrated a construction technique for
large-scale, single-shell Metglas magnetic shields, consisting of
little material (thicknesses of only 102--122 $\mu$m), which yields a
high DC axial shielding factor in the Earth's magnetic field.  We
emphasize that we obtained these results in a passive DC environment,
without the use of magnetic shaking or attenuation with any other
magnetic coils.  We believe that the results we obtained are the
result of our careful construction technique, which minimizes the
impact of any possible gaps between adjacent foil windings and layers.

\begin{acknowledgments}
This work was supported in part by the U.\ S.\ Department of Energy
Office of Nuclear Physics under Award Number DE-FG02-08ER41557, and by
the University of Kentucky.  We thank R.\ Golub and A.\ P\'{e}rez
Galv\'{a}n for valuable discussions.  B.P.\ thanks the Kellogg
Radiation Laboratory at the California Institute of Technology for the
hospitality during a visit during which time this article was written.
\end{acknowledgments}

\end{document}